\documentclass[showpacs,amsmath,amssymb,prb,superscriptaddress,twocolumn]{revtex4}
\usepackage[T1]{fontenc}
\usepackage{amssymb,amsmath}
\usepackage{times}
\usepackage{graphicx}
\usepackage{wasysym}
\usepackage{subfigure}
\usepackage{bm}

\begin{document}

\title{Self-consistent calculation of metamaterials with gain}

\author{A.~Fang}
\affiliation{Department of Physics and Astronomy and Ames Laboratory,
             Iowa State University, Ames, Iowa 50011, U.S.A.}

\author{Th.~Koschny}
\affiliation{Department of Physics and Astronomy and Ames Laboratory,
             Iowa State University, Ames, Iowa 50011, U.S.A.}
\affiliation{Department of Materials Science and Technology and Institute of Electronic Structure and Laser, FORTH,
              University of Crete, 71110 Heraklion, Crete, Greece}

\author{M.~Wegener}
\affiliation{Institut f\"ur Angewandte Physik and DFG-Center for Functional Nanostructures (CFN),
             Universit\"at Karlsruhe (TH), D-76128 Karlsruhe, Germany}

\author{C.~M.~Soukoulis}
\affiliation{Department of Physics and Astronomy and Ames Laboratory,
             Iowa State University, Ames, Iowa 50011, U.S.A.}
\affiliation{Department of Materials Science and Technology and Institute of Electronic Structure and Laser, FORTH,
              University of Crete, 71110 Heraklion, Crete, Greece}

\date{\today}


\begin{abstract}
We present a computational scheme allowing for a self-consistent treatment of a
dispersive metallic photonic metamaterial coupled to a gain material
incorporated into the nanostructure.  The gain is described by a generic
four-level system.  A critical pumping rate exists for compensating the loss of
the metamaterial.  Nonlinearities arise due to gain depletion beyond a certain
critical strength of a test field.  Transmission, reflection, and absorption
data as well as the retrieved effective parameters are presented for a lattice
of resonant square cylinders embedded in layers of gain material and split
ring resonators with gain material embedded into the gaps.
\end{abstract}



\maketitle

The field of metamaterials \cite {1,2} is driven by fascinating and
far-reaching theoretical visions such as, e.g., perfect lenses, \cite{3} 
invisibility cloaking, \cite{4,5} and enhanced optical nonlinearities. \cite{6}
This emerging field has seen spectacular experimental progress in recent years. \cite{1,2}   
Yet, losses are orders of magnitude too large for the envisioned applications. 
Achieving such reduction by further design optimization appears to be out of reach. 
Thus, incorporation of active media (gain) might come as a cure.
The dream would be to simply inject an electrical current into the active
medium, leading to gain and hence to compensation of the losses. 
However, experiments on such intricate active nanostructures do need guidance
by theory via self-consistent calculations (using the semi-classical theory of
lasing) for realistic gain materials that can be incorporated into or close to
dispersive media to reduce the losses at THz or optical frequencies.  
The need for \textit{self-consistent} calculations stems from the
fact that increasing the gain in the metamaterial, the metamaterial properties
change, in turn changes the coupling to the gain medium until a steady-state
is reached. 
A specific geometry to overcome the severe loss problem of optical
metamaterials and to enable bulk metamaterials with negative magnetic and
electric response and controllable dispersion at optical frequencies is to
interleave active optically pumped gain material layers with the passive
metamaterial lattice.

For reference, the best fabricated negative-index material operating at around
$1.4\,\mathrm{\mu m}$ wavelength \cite{7} has shown a figure of merit 
$(\mathrm{FOM}) = -\mathrm{Re}(n)/\mathrm{Im}(n) \approx 3$, where $n$ is the
effective refractive index.  This experimental result is equivalent to an
absolute absorption coefficient of $\alpha=3\times10^4 \,\mathrm{cm}^{-1}$,
which is even larger than the absorption of typical direct-gap semiconductors
such as, e.g., GaAs (where $\alpha=10^4 \,\mathrm{cm}^{-1}$).  So it looks
difficult to compensate the losses with this simple type of analysis, which
assumes that the bulk gain coefficient is needed. However, the effective gain
coefficient, derived from self-consistent microscopic calculations, is a more appropriate
measure of the combined system of metamaterial and gain.  Due to pronounced
local-field enhancement effects in the spatial vicinity of the dispersive
metamaterial, the effective gain coefficient can be substantially larger than
its bulk counterpart. While early models using simplified gain-mechanisms such as
explicitly forcing negative imaginary parts of the local gain material's
response function produce unrealistic strictly linear gain, our self-consistent
approach presented below allows for determining the range of parameters for which
one can realistically expect linear amplification and linear loss compensation
in the metamaterial.  
To fully understand the coupled metamaterial-gain system, we have
to deal with time-dependent wave equations in metamaterial systems by coupling
Maxwell's equations with the rate equations of electron populations describing
a multi-level gain system in semi-classical theory. \cite{8}

In this paper, we apply a detailed computational model to the problem of
metamaterials with gain.  The generic four-level atomic system tracks fields
and occupation numbers at each point in space, taking into account energy exchange
between atoms and fields, electronic pumping and non-radiative decays. \cite{8} 
An external mechanism pumps electrons from the ground state level $N_0$ to the
third level $N_3$ at a certain pumping rate $\Gamma_\mathrm{pump}$, which is
proportional to the optical pumping intensity in an experiment. 
After a short lifetime $\tau_{32}$ electrons transfer non-radiatively into the
metastable second level $N_2$.  The second level ($N_2$) and the first level
($N_1$) are called the upper and lower lasing levels. Electrons can be
transferred from the upper to the lower lasing level by spontaneous and
stimulated emission. At last, electrons transfer quickly and non-radiatively
from the first level ($N_1$) to the ground state level ($N_0$). The lifetimes
and energies of the upper and lower lasing levels are $\tau_{21},\ E_2$ and
$\tau_{10},\ E_1$, respectively.  The center frequency of the radiation is
$\omega_a=(E_2-E_1)/\hbar$ which is chosen to equal $2\pi \times 10^{14}
\,\mathrm{Hz}$.  The parameters $\tau_{32}$, $\tau_{21}$, and $\tau_{10}$ are
chosen $5\times10^{-14}$, $5\times10^{-12}$, and $5\times10^{-14}\,\mathrm{s}$, respectively.  The total electron density,
$N_0(t=0) = N_0(t) + N_1(t) + N_2(t) + N_3(t) = 5.0\times 10^{23}\,\mathrm
{/m^3}$, and the pump rate $\Gamma_\mathrm{pump}$ are controlled variables
according to the experiment.
The time-dependent Maxwell equations are given by $\nabla\times \mathbf {E} = -\partial
 \mathbf {B}/\partial t$ and $\nabla\times \mathbf {H} = \varepsilon \varepsilon_o \partial
 \mathbf {E}/\partial t + \partial \mathbf {P}/\partial t$, where $\mathbf {B}=\mu\mu_o \mathbf {H}$ and
$\mathbf {P}$ is the dispersive electric polarization density from which the
amplification and gain can be obtained.
Following the single electron case, we can show \cite{8} that the polarization
density $\mathbf {P}(\mathbf {r},t)$ in the presence of an electric field obeys locally the following
equation of motion
\begin{equation}
 \label{Eqn:1}
 \frac{\partial^2 \mathbf {P}(t)}{\partial t^2} +
 \Gamma_a\frac{\partial \mathbf {P}(t)}{\partial t} +
 \omega_a^2 \mathbf {P}(t) \ =\
 -\sigma_a \Delta N(t) \mathbf {E}(t)
\end{equation}
where $\Gamma_a$ is the linewidth of the atomic transition
$\omega_a$ and is equal to 
$2\pi \times 5 \times 10^{12} \,\mathrm{Hz}$ or 
$2\pi \times 20 \times 10^{12} \,\mathrm{Hz}$. The
factor $\Delta N(\mathbf {r},t) = N_2(\mathbf {r},t) - N_1(\mathbf {r},t)$ is the population inversion that
drives the polarization, and $\sigma_a$ is the coupling strength of $\mathbf {P}$ to the
external electric field and its value is taken to be $10^{-4}\,\mathrm
{C^2/kg}$. 
It follows \cite{8} from Eqn.~\ref{Eqn:1} that the
amplification line shape is Lorentzian and homogeneously broadened.\cite {8a} 
The occupation numbers at each spatial point vary according to
\begin{subequations} 
 \begin{align}
 \frac{\partial N_3}{\partial t} &= \Gamma_\mathrm{pump}\, N_0 -
                                    \frac{N_3}{\tau_{32}} \\
 \frac{\partial N_2}{\partial t} &= \frac{N_3}{\tau_{32}} +
                                    \frac{1}{\hbar\omega_a} \mathbf {E}\cdot\frac{\partial \mathbf {P}}{\partial t}  -
                                    \frac{N_2}{\tau_{21}} \\
 \frac{\partial N_1}{\partial t} &= \frac{N_2}{\tau_{21}} -
                                    \frac{1}{\hbar\omega_a} \mathbf {E}\cdot\frac{\partial \mathbf {P}}{\partial t}  -
                                    \frac{N_1}{\tau_{10}}  \\
 \frac{\partial N_0}{\partial t} &= \frac{N_1}{\tau_{10}} -
                                    \Gamma_\mathrm{pump}\, N_0 
 \end{align}
\end{subequations}
where $\frac{1}{\hbar\omega_a}\mathbf {E}\cdot\frac{\partial \mathbf {P}}{\partial t}$ 
is the induced radiation rate or excitation rate depending on its sign.

\begin{figure}
 \centerline{\includegraphics[angle=-90, width=8.cm]{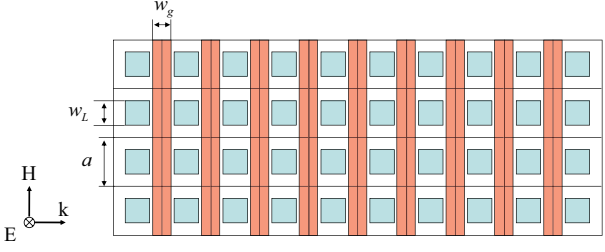}}
 \caption{%
 (Color online)
 Square lattice of dielectric square cylinders (blue) that have a Lorentz behavior
 embedded in layers of gain material (red).
 The dielectric constant of the cylinders is given by
 $\varepsilon = 1 + \omega_p^2/(\omega_p^2 -2i\omega\gamma -\omega^2)$,
 where $f_p = \omega_p/2\pi = 100 \,\mathrm{THz}$ and $\gamma = 2\pi f$,
 and $f$ takes different values in the cases we have examined. The dimensions are
 $a=80\,\mathrm {nm}$, $w_{L}=40\,\mathrm {nm}$ and $w_g=30\,\mathrm {nm}$.
 }
 \label{fig:1}
\end{figure}

In order to solve the behavior of the active materials in the electromagnetic
fields numerically, the finite-difference time-domain (FDTD) technique is
utilized, \cite{9} using an approach similar to the one outlined in Refs. 10--12.  
In the FDTD calculations, 
the discrete time and space steps are chosen to be $\Delta t=8.33 \times
10^{-18}\,\mathrm{s}$ and $\Delta x=5.0\times10^{-9}\,\mathrm{m}$ for
simulations on the structure as shown in Fig.~1, and $\Delta t=8.33 \times
10^{-19}\,\mathrm{s}$ and $\Delta x=1.0\times10^{-9}\,\mathrm{m}$ for
simulations on the structure as shown in Fig.~5.  The initial condition is that
all the electrons are in the ground state, so there is no field, no
polarization and no spontaneous emission. Then the electrons are pumped from
$N_0$ to $N_3$ (then relaxing to $N_2$) with a constant pumping rate $\Gamma_\mathrm{pump}$. The
system begins to evolve according to the system of equations above.

\begin{figure}
 \centering
  \includegraphics[width=0.35\textwidth]{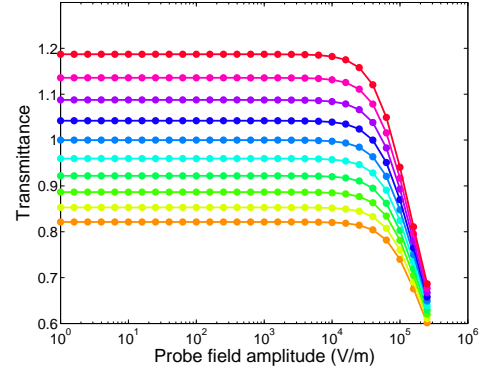}
 \caption{%
 (Color online)
 The transmittance vs.\@ probe field amplitude for the loss-compensated metamaterial of Fig.~1 with gain bandwidth $5\,\mathrm {THz}$, 
 loss bandwidth $20\,\mathrm {THz}$ (i.e., $f = 10\,\mathrm {THz}$), for different pumping rates $\Gamma_\mathrm{pump}$. $\Gamma_\mathrm{pump}$ 
 is increased from $2.15\times 10^9\,\mathrm {s^{-1}}$ (lowest) to $3.05\times 10^9\,\mathrm {s^{-1}}$ (highest) in steps of $0.1\times 10^9\,\mathrm {s^{-1}}$. 
 The metamaterial response is linear in a very wide range. When the loss-compensated
 transmission is exactly unity, the pumping rate $\Gamma_\mathrm{pump}=2.65\times 10^9\,\mathrm {s^{-1}}$, which is called the critical pumping rate. 
 For incident fields stronger than $10^4\,\mathrm{V/m}$ this metamaterial becomes non-linear.
 }
\label{fig:2}
\end{figure}

We have performed numerical simulations on one-dimensional (1D) and two-dimensional (2D) systems with gain. \cite {13} 
Previous studies \cite{14,15,16,17,18} have considered loss reduction by incorporating gain
but where not self-consistent (see introduction). \cite{14,15,16,17} 
As the first simple model system, we will discuss a 2D metamaterial system (shown in Fig.~1) which consists of layers 
of gain material and dielectric wires that have a resonant Lorentz type electric response to emulate the resonant elements
in a realistic metamaterial.  
We will have to study whether we will be able to compensate the losses of the metamaterials associated with the Lorentz 
resonance in the wires by the amplification provided by the gain material layers without destroying the linear response
of the metamaterial.  
First we generate a narrow band Gaussian pulse of a given amplitude and let it propagate through the metamaterial without gain, 
and we calculate the transmitted signal emerging from the metamaterial which has also Gaussian profile but the amplitude 
is much smaller than that of the incident pulse depending on how much loss occurs in the metamaterial. 
Then we introduce the gain and start increasing the pumping rate and find a critical pumping rate, $\Gamma_\mathrm{pump}=2.65\times 10^9\,\mathrm {s^{-1}}$, for which the transmitted pulse 
is of the same amplitude as the incident pulse.  
In addition, for fixed pumping rate, we start increasing the amplitude of the incident Gaussian pulse and we would like to 
see how high we can go in the strength of the incident electric field and still have full compensation of the losses,
i.e. the transmitted signal equals the incident signal, independent on the signal strength.
In this region we have compensated loss and still linear response of the metamaterial; 
here, the shape of the transmitted Gaussian is only affected by the dispersion but not dependent on the signal strength.
\begin{figure}
 \centering
  \includegraphics[width=0.35\textwidth]{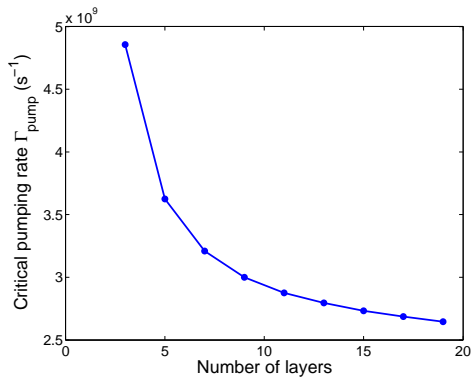}
 \caption{%
  (Color online)
  The critical pumping rates for different numbers of layers of the system in Fig.~1. Parameters for gain and dielectric materials are the same as Fig.~2.
  }
  \label{fig:4}
\end{figure}

We have calculated the transmission versus the strength of the electric field of the incident signal for several pumping rates close to the critical pumping rate. 
As shown in Fig.~2, we found that for a rather broad region of low intensity input signal we have a linear response 
all the way up to incident electric field of $10^3\,\mathrm{V/m}$. If we use only three layers (rods - gain material - rods), 
 the critical pumping rate is $4.85\times 10^9\, \mathrm {s^{-1}}$, which is two times higher than the 19-layer case of Fig.~1. In Fig.~3, 
we present detailed results for the critical pumping rate versus the number of layers of the system shown in Fig.~1. Notice that as the number of 
layers increases, the critical $\Gamma_{\mathrm {pump}}$ decreases.
 The linear regime for three layers exists up to $10^4\, \mathrm {V/m}$, and for higher strength drops slower than that of Fig.~2. 
 In all the following simulations, the strength of 
the incident signal is chosen to be $10\, \mathrm {V/m}$, which is far away from $10^3\,\mathrm{V/m}$, so we operate in the 
linear regime of the metamaterial.
As an example, we have studied three layers, rods - gain material - rods, 
to see how much $\Gamma_\mathrm{pump}$ we need to compensate the losses. 
As expected, we found that $\Gamma_\mathrm{pump}$ is proportional to the imaginary part of the 
permittivity $\varepsilon$ of the dielectric. 

\begin{figure}
 \centering
  \includegraphics[width=0.35\textwidth]{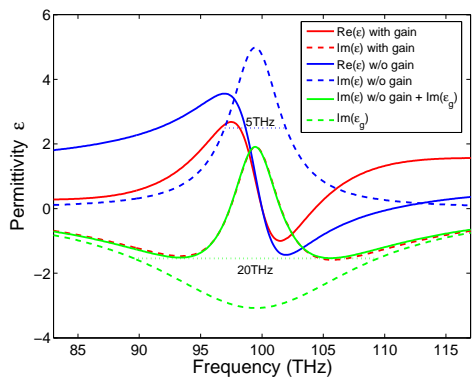}
 \caption{%
  (Color online)
  The retrieved results for the real and the imaginary parts of the effective
  permittivity, $\varepsilon$, with and without gain.
  In addition, we have plotted $\mathrm{Im}(\varepsilon_g)$ versus frequency.
  Below compensation, $t=0.75$; 
  Gain and Lorentz bandwidths are  $20$ and $5\,\mathrm {THz}$, respectively.
 }
 \label{fig:4}
\end{figure}

We first present results for three layers of the system shown in Fig.~1. 
First, the full width at half maximum (FWHM) for Lorentz dielectric and gain are chosen to be $5$ and $20\,\mathrm{THz}$, 
respectively.
With the introduction of gain the absorption at the resonance 
frequency of $100 \,\mathrm{THz}$ decreases, ultimately reaching zero (not shown).  
So the gain compensates the losses.  
In Fig.~4, we plot the retrieved results for the real and imaginary parts of $\varepsilon$ 
without gain and with gain slightly below compensation (see Ref. 19 for the retrieval method).
Notice that we can have the $\mathrm {Re}(\varepsilon) \approx -1$ with $\mathrm {Im}(\varepsilon)
\approx 0$ at $102\,\mathrm {THz}$, slightly off the resonance frequency.
From Fig.~4, one can also see that $\mathrm {Re}(\varepsilon) \approx 2.5$ 
with  $\mathrm {Im}(\varepsilon) \approx 0$ at $97\,\mathrm {THz}$. 
So one can obtain a lossless metamaterial with positive or negative $\mathrm {Re}(\varepsilon)$.
Once we introduce gain, the imaginary part of $\varepsilon$ of our total system
with gain is equal to the sum of $\mathrm{Im}(\varepsilon)$ without gain
and the imaginary part of $\varepsilon_g$, the dielectric function of the gain
material. 
This result is unexpected, because there is no coupling between the 2D Lorentz dielectric with the gain material. 
This is indeed true because of the continuous shape of the Lorentz dielectric cylinders and the gain material slabs have zero 
depolarization field. In contrast to finite length wires (hence a 3D problem) where the dipole interactions between Lorentz dielectric 
and gain material would be dominated by the quasi-static nearfield $\it {O}(1/r^3)$, here the interaction is order $\it {O}(\omega \ln|kr|)$, 
only via the propagating field, and much weaker. 
Therefore, for this 2D model, gain and loss are approximately independent.
The behavior would obviously be different in a 3D situation, which, however, is computationally excessively demanding.
Thus, we consider
a 2D version of the split ring resonator (SRR) as a more realistic and also more relevant model. 
Here, the relevant polarization is across the finite SRR gap and, therefore,
the coupling to the gain material is in fact dipole like. 

\begin{figure}
 \centerline{\includegraphics[angle=-90, width=0.4\textwidth]{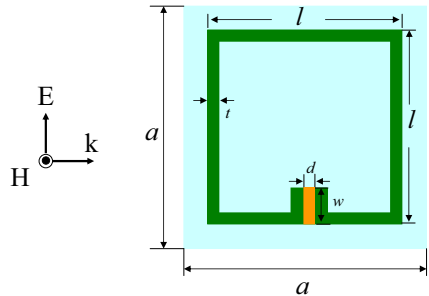}}
 \caption{%
  (Color online)
  Geometry for a unit cell of the square SRR system with gain.
  The gain (shown in orange) is introduced in the gap region of the SRR.
  The dimensions are 
  $a=100\,\mathrm {nm}$, $l=80\,\mathrm {nm}$, $t=5\,\mathrm {nm}$,
  $d=4\,\mathrm {nm}$ and $w=15\,\mathrm {nm}$.
 }
 \label{fig:5}
\end{figure}

\begin{figure}
 \centering
  \subfigure{
   \includegraphics[width=0.35\textwidth]{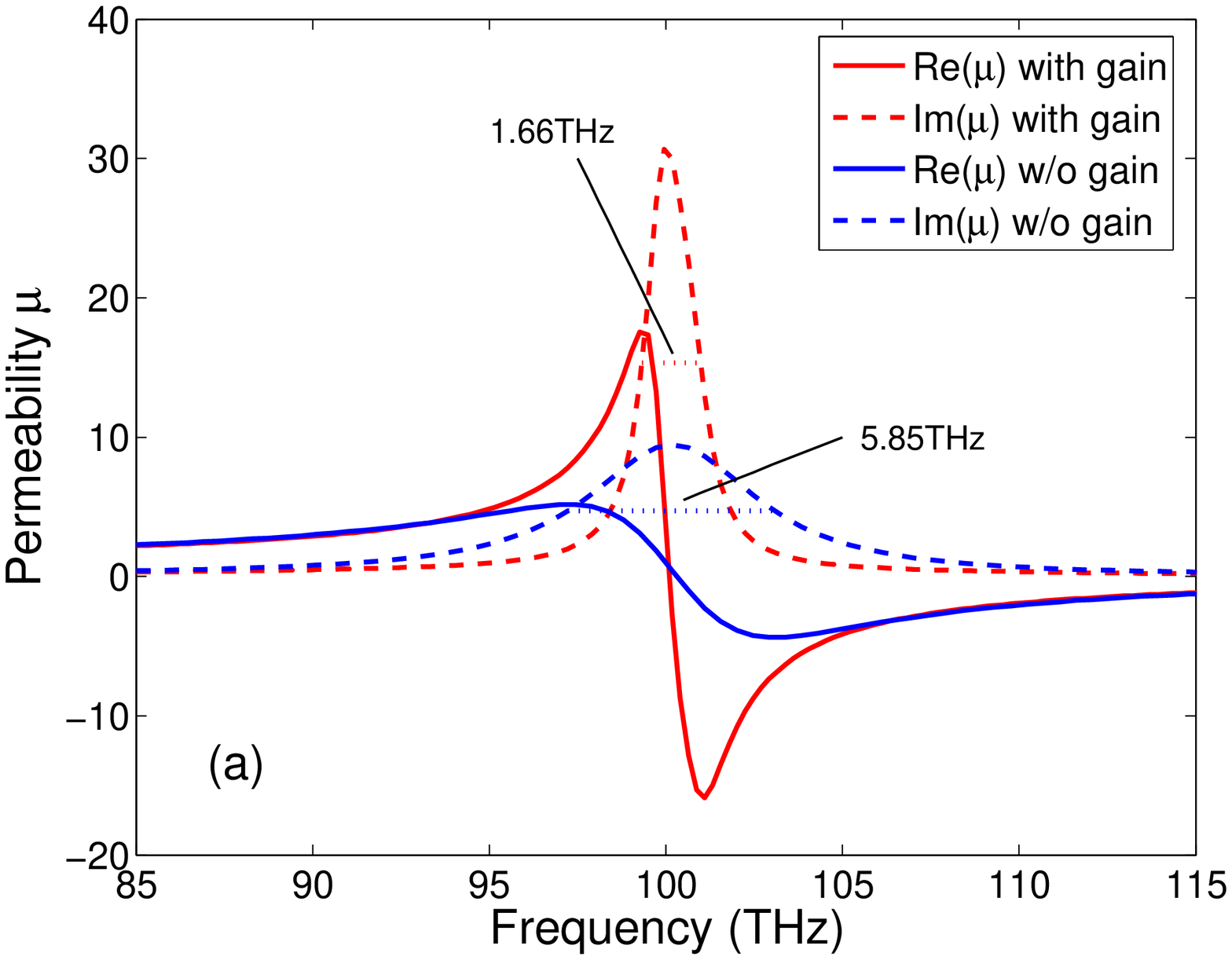}
   }
 \centering
  \subfigure{
   \includegraphics[width=0.35\textwidth]{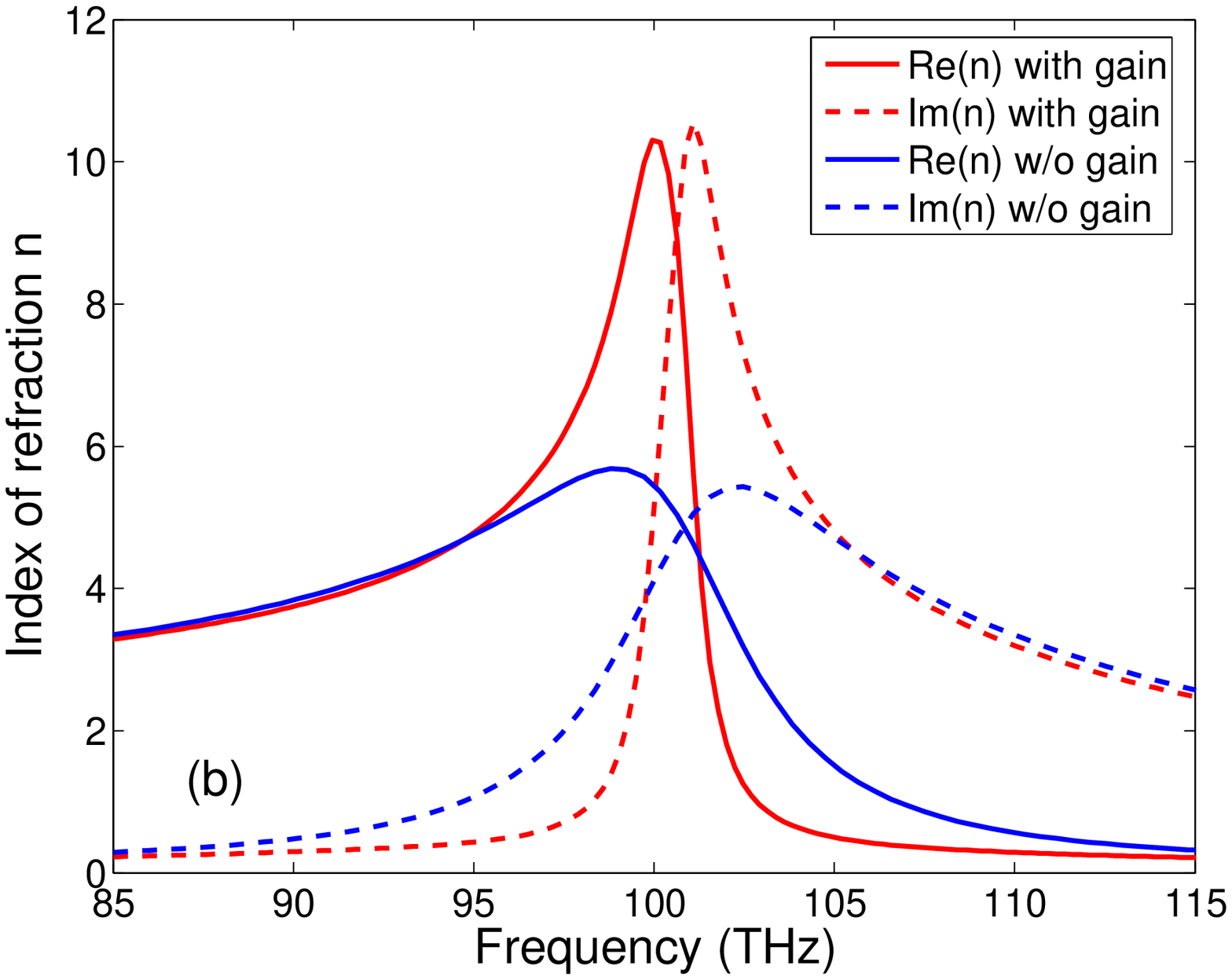}
  }
 \caption{%
  (Color online)
  The retrieved results for the real and the imaginary parts of
  (a) the effective permeability $\mu$ and
  (b) the corresponding effective index of refraction $n$,
  with and without gain
  for a pumping rate $\Gamma_\mathrm{pump}=1.4\times 10^9\,\mathrm {s^{-1}}$.
  The gain bandwidth is $20\,\mathrm {THz}$.
  Notice that the width of the resonance with gain is $1.66\,\mathrm {THz}$.
 }
 \label{fig:6}
\end{figure}

In Fig.~5, we present the unit cell of our SRR system with gain material embedded in the SRR gap. 
The dimensions of the SRR are chosen such that a magnetic resonance frequency at $100\,\mathrm {THz}$ results, 
which can overlap with the peak of the emission of the gain material. 
The FWHM of the gain material is $20\,\mathrm {THz}$, and $\Gamma_\mathrm{pump}$ is $1.4\times 10^9\,\mathrm {s^{-1}}$. 
 Simulations are done for one layer of the square SRR. 
 In Fig.~6a, we plot the retrieved results of the real and the imaginary parts of the magnetic permeability, $\mu$, with and without gain.
With the introduction of gain, the weak and broad resonant effective $\mu$ (FWHM = $5.85\,\mathrm {THz}$) of the lossy SRR becomes strong and narrow (FWHM = $1.66\,\mathrm {THz}$);
the gain effectively undamps the LCR resonance of the SRR.
Notice that here losses in the magnetic effective response are compensated by electric gain in the SRR gap.
So with the introduction of gain, we obtain a negative $\mu$ with a very small
imaginary part in an otherwise typical SRR response, 
which means that the losses have been compensated by the gain. 
In Fig.~6b, we plot the retrieved results for the effective index of refraction $n$, with and without gain.
Note that for a lossless SRR $n$ is purely real away from the resonance and imaginary in a small band above the resonance
where $\mu$ is negative.
Comparing $\mathrm {Re}(n)$ slightly below the resonance at $97\,\mathrm {THz}$, 
we find an effective extinction coefficient 
$\alpha = (\omega/c)\,\mathrm{Im}(n) \approx 3.50\times 10^4\,\mathrm{cm}^{-1}$ without gain, 
and $\alpha\approx 1.24\times 10^4\,\mathrm{cm}^{-1}$ with gain, and  
hence an effective amplification of $\alpha\approx -2.26\times 10^4\,\mathrm{cm}^{-1}$.
This is much larger than the expected amplification 
$\alpha \approx -1.39\times 10^3\,\mathrm{cm}^{-1}$ for the gain material at the given pumping rate. \cite {20} 
The difference can be explained by the field enhancement in the gap of the resonant SRR.
The induced electric field in the gap is around $550\, \mathrm {V/m}$, which is still in the linear regime, and
 the incident electric field is $10\, \mathrm {V/m}$. Indeed, taking the observed field enhancement factor in the SRR gap of $\approx 55$, 
the energy per unit cell produced by the gain material inside the gap is $\approx 18$ 
times larger than for the homogeneous gain medium 
which compares very well to the factor $\approx 20$ between the simulated SRR effective 
medium and the homogeneous gain medium.
If we further increase the pumping rate the magnetic resonance becomes even narrower 
($0.96\,\mathrm{THz}$ for $\Gamma_\mathrm{pump}=1.8\times 10^9\,\mathrm {s^{-1}}$).
When the pumping rate reaches $\Gamma_\mathrm{pump}=1.9\times 10^9\,\mathrm {s^{-1}}$, $\mathrm{Im}(\mu)$ becomes negative
and we have overcompensated at the resonance frequency.
By increasing $\Gamma_\mathrm{pump}$ even more ($\approx 5\times 10^9\,\mathrm {s^{-1}}$) 
one starts seeing lasing (spasing) \cite{21,22} in our system (not shown), 
which is not in the focus of this work. As long as we are in the linear regime, we do not need to have a self-consistent calculation, our results
 agree very well with the results obtained using the susceptibilities given in Ref.~9. However, the self-consistent calculation is 
necessary to determine the range of signals for which we can expect approximately linear response and it is needed 
if we have very strong fields and we are in the nonlinear regime, especially when we want to study lasing.

In conclusion, we have proposed and numerically solved a self-consistent model
incorporating gain in 2D dispersive metamaterials.  We show numerically that
one can compensate the losses of the dispersive metamaterials.  There is a
relatively wide range of signal amplitudes for which the loss-compensated
metamaterial still behaves linearly; at higher amplitudes the response is
non-linear due to the gain.
As an example, we have demonstrated that the losses of the magnetic
susceptibility $\mu$ of the SRR can be easily compensated by the gain material.
The pumping rate needed to compensate the loss is much smaller than the bulk
gain.
This aspect is due to the strong local-field enhancement inside the SRR gap. 


Work at Ames Laboratory was supported by the Department of Energy (Basic Energy
Sciences) under Contract No.~DE-AC02-07CH11358.
This work was partially supported by the European Community FET project PHOME
(Contract No.~213390) and the Office of Naval Research (Award
No.~N00014-07-1-0359).

\bibliographystyle{apsrev}

\begin{thebibliography}{99}
\bibitem{8a}
The real and imaginary parts of the gain profile is given by
$P/\varepsilon_0 E = \chi'+i\chi''$, where 
$\chi' = -\chi_0'' \Delta x/(1+\Delta x^2)$ and 
$\chi'' = \chi_0'' /(1+\Delta x^2)$ with
$\Delta x = 2(\omega-\omega_a)/\Gamma_a$ and
$\chi_0''= -\sigma_a \Delta N/(\varepsilon_0^{}\omega_a\Gamma_a)$ where $\Delta N = N_2-N_1$.
\bibitem{13}
We first check that a well-defined lasing threshold exists for a gain material
slab with a width of $500\,\mathrm {nm}$. The pumping rate for lasing threshold
is $1.6\times 10^9\,\mathrm {s^{-1}}$, which corresponds to incident power of
$65\,\mathrm {W/mm^2}$ and the output power is $4\times 10^{-3}\,\mathrm
{W/mm^2}$. If the pumping rate is much higher than the lasing threshold, i.e.
$3.5\times 10^9\,\mathrm {s^{-1}}$, the occupation numbers $N_2$ and $N_1$
oscillate as a function of time and finally saturate to the stable values. For
the parameters used, $N_2 \approx 7.5\times 10^{-3}~N_0$ and $N_1\approx
2.0\times 10^{-4}~N_0$.
\bibitem{20}
The extinction coefficient (amplification coefficient) of the gain material can be calculated from 
$\alpha=\omega\chi''/(2c\sqrt{\varepsilon_r})$ given in chapter 7 of Ref.~8.
For the pumping rate $\Gamma_\mathrm{pump}=1.4\times 10^9\,\mathrm {s^{-1}}$ and $f=100\,\mathrm{THz}$ 
we obtain $\Delta N=6.9\times 10^{-3}N_i$ ($N_i = N_o(t=0)=5.0\times 10^{23} \, \mathrm {/m^3}$), 
and hence $\alpha=-1.50\times 10^3\mathrm{cm}^{-1}$ at the resonance. 
In our case for $97\,\mathrm{THz}$, we get $\alpha=-1.39\times 10^3\mathrm{cm}^{-1}$.
\end{thebibliography}

\end{document}